\documentclass[%
 aip,
 amsmath,amssymb,
 reprint,%
]{revtex4-1}
\usepackage{graphicx}
\usepackage{color}

\begin{document}

\title{Stochastic Magnetoelectric Neuron for Temporal Information Encoding}%

\author{Kezhou Yang}
\author{Abhronil Sengupta}
\email{sengupta@psu.edu}

\affiliation{{\small Electrical Engineering \& Materials Science and Engineering, \\
The Pennsylvania State University, University Park, PA 16802, USA}}%

\begin{abstract}
{\small
Emulating various facets of computing principles of the brain can potentially lead to the development of neuro-computers that are able to exhibit brain-like cognitive capabilities. In this letter, we propose a magnetoelectronic neuron that utilizes noise as a computing resource and is able to encode information over time through the independent control of external voltage signals. We extensively characterize the device operation using simulations and demonstrate its suitability for neuromorphic computing platforms performing temporal information encoding.  }
\end{abstract}

\maketitle
As a possible route to reduce the huge computational overhead of deep learning based Artificial Intelligence platforms, ``brain-like" algorithmic primitives (referred to as Spiking Neural Networks, SNNs, herein) and hardware are being actively explored due to its promise of enabling low-power, event-driven asynchronous neuromorphic hardware \cite{monroe2014neuromorphic,merolla2014million,davies2018loihi}. However, emerging post-CMOS technology \cite{sengupta2015spin,jackson2013nanoscale,kuzum2011nanoelectronic} based neuromorphic computing faces significant challenges currently. From the algorithmic perspective, SNN computing models are a significant shift from the traditional deep non-spiking networks (current de-facto standard) due to the additional time domain encoding of information. Hence, classification accuracies provided by such networks are still limited than their non-spiking counterparts \cite{sengupta2018going}. Further, it is unclear whether emerging neuromorphic devices based on spintronics, resistive memories, phase-change memories  would still exhibit multi-level characteristics at aggressively scaled device dimensions (the key characteristic being leveraged in these non-volatile devices). Additionally, such devices are characterized by enhanced stochasticity during the switching process. As a pathway to overcome these limitations, we formulate our solution against two complementary backdrops by noting that any neuromorphic computing platform operates on the basis of two fundamental principles -- how information is encoded in the time-varying spike train and how computing occurs to generate the spike train itself.

Inferring information from the time-domain spiking behavior of neurons in an SNN are either done in a rate-based fashion or temporally. In rate-based networks, information is considered to be encoded in the total number of spikes generated by a neuron in a sufficiently long time-window. However, in temporal-encoding, the precise timing of the spikes is believed to carry information. The additional information encoding capacity in the temporal spike encoding has the potential to scale up the recognition performance of SNNs \cite{liu2018pt,liu2017mt,masquelier2007unsupervised,wang2017spiketemp,bagheri2018training}. Completely agnostic to the manner of information encoding in the network, is the nature of computing in the neural nodes. Standard deterministic deep learning frameworks enabled by spintronic devices and other post-CMOS technologies have been explored. In such scenarios, device-level non-idealities are usually treated as a disadvantage. More recently, stochasticity inherent in such devices have been exploited for computing to implement stochastic neural computing \cite{sengupta2016probabilistic}. From a brain-emulation perspective, there is increasing evidence that the brain performs probabilistic computation through its noisy neurons, synapses and dendrites \cite{nessler2013bayesian}. Theoretical understanding for benefits (for instance, training convergence) of stochastic computation in neurons \cite{nessler2013bayesian} and synapses \cite{neftci2017stochastic} have also started in earnest. Such a stochastic computing framework enables state-compressed neural networks (implemented using single-bit scaled binary magnetic devices) where the accuracy loss due to bit-compression is compensated by the the additional probabilistic encoding of information.  

\begin{figure}
\centering
\includegraphics[width=0.38\textwidth]{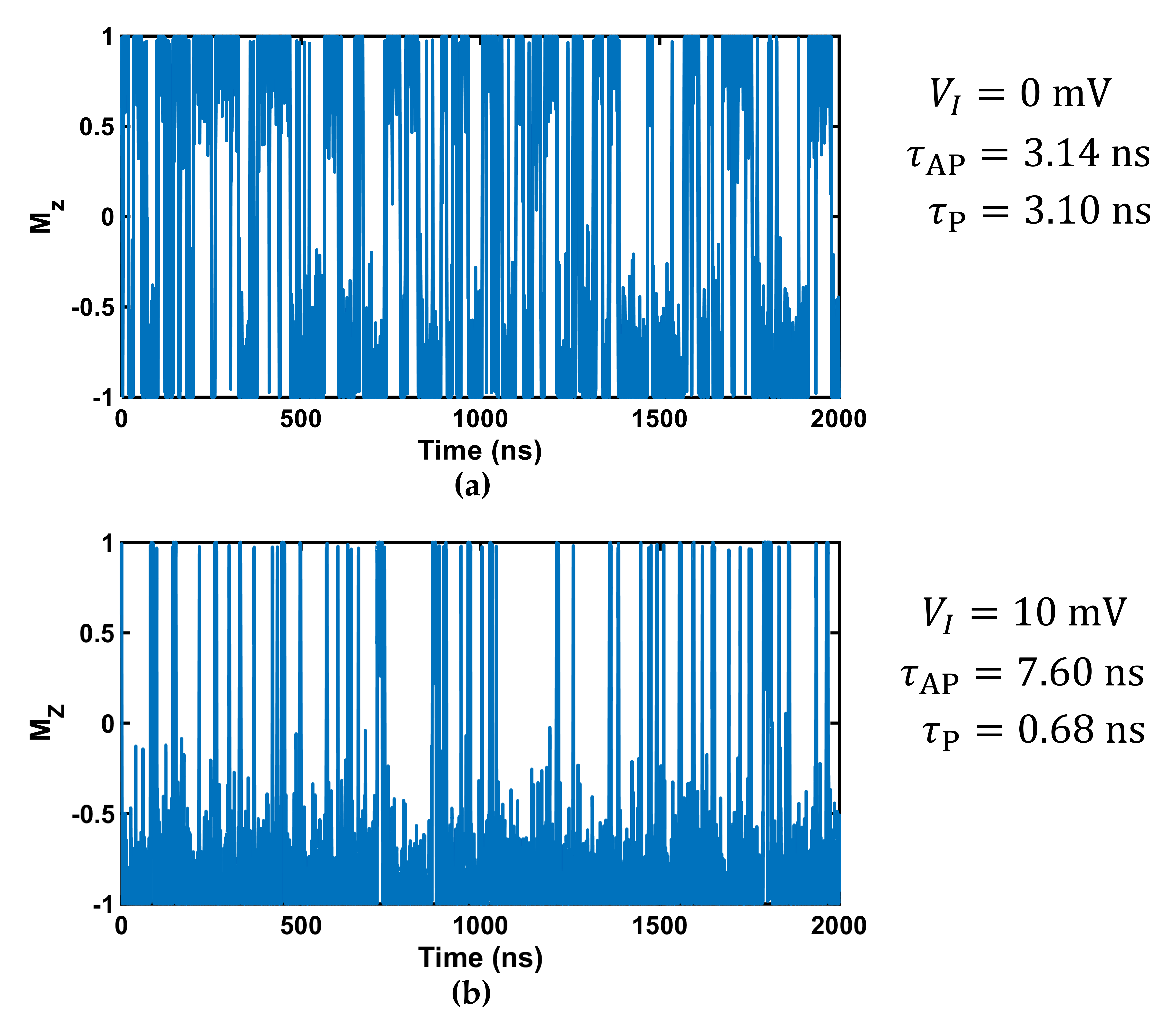}
\caption{\scriptsize{Temporal switching characteristics of the $z$-component of magnetization are shown for a $2k_{B}T$ barrier height magnet. With change in $V_I$, both the device lifetimes $\tau_{AP}$ and $\tau_P$ gets modulated, thereby varying the device firing rate.}}
\label{fig1}
\vspace{-2mm}
\end{figure}

Most of the current work on utilizing magnetic devices as stochastic neurons rely on rate encoding models where the rate of spiking of a neuron is a non-linear function of the weighted summation of synaptic inputs. Such a functionality can be directly mapped to the switching behavior of a Magnetic Tunnel Junction (MTJ) \cite{julliere1975tunneling} in the presence of thermal noise at non-zero temperatures \cite{scholz2001micromagnetic}. An MTJ consists of two nanomagnets sandwiching a spacer layer (typically an oxide such as MgO). While the magnetization of one of the layers is magnetostatically ``pinned" or ``hardened" in a particular direction, the magnetization of the other layer can be switched by an external stimulus, such as a spin current or magnetic field. The two layers are referred to as the ``Pinned" layer (PL) and ``Free" layer (FL) respectively. Depending on the relative orientation of the two magnets, the device exhibits a high-resistance anti-parallel (AP) state (when the magnetizations of the two layers have opposite direction) and a low-resistance parallel (P) state (when the magnetizations of the two layers have the same direction). These two states of the magnet are stabilized by an energy barrier determined by the anisotropy and volume of the magnet. As the barrier height is scaled down, the magnet undergoes spontaneous random telegraphic switching between the two stable states. Fig. \ref{fig1} depicts the temporal magnetization dynamics of a $\sim 2k_{B}T$ ($k_B$ is Boltzmann constant and $T$ is absolute temperature) barrier height magnet. The magnet resides in the P and AP states with characteristic lifetimes $\tau_{P}$ and $\tau_{AP}$. The lifetime of the device in each state can be controlled by the magnitude or direction of an external current flowing through the magnetic stack \cite{slonczewski1989conductance}. At zero bias current, the lifetimes are equal and determined by the magnet barrier-height. Note that this is a first-order modelling. In practical device implementation, $50\%$ switching probability may not be achieved exactly at zero bias current due to the presence of device imperfections, stray fields, and other non-idealities. With the application of an external ``write" current, the magnitude of the firing rate of the neuron, $\tau = \frac{\tau_{AP}}{\tau_{P}+\tau_{AP}}$ gets modulated. The rate of spiking of the neuron varies in a non-linear sigmoid fashion with respect to the input current \cite{liyanagedera2017stochastic}.

As mentioned previously, the vast majority of works, utilizing SNNs as a computational paradigm, have relied on rate based information encoding. Essentially, an averaging process is performed wherein the total number of spikes generated by a particular neuron is counted over a specific timing window. This rate is used for inference purposes as well. Recently, temporal information encoding is being actively explored in the domain of neuromorphic algorithms \cite{liu2018pt,liu2017mt,masquelier2007unsupervised,wang2017spiketemp,bagheri2018training} while it has been long studied in the domain of stochastic non-linear systems \cite{sinha1998dynamics,kohar2014enhanced,kohar2017implementing}. The principal benefits of using temporal encoding for modelling spiking behavior are multiple. Since information is now embodied in specific spike timings instead of the signal rate (which needs to be observed over longer periods of time), such neural codes can be sparse and much faster to avoid temporal-averaging effect. Sparsity in neural spikes will translate to huge benefits in neuromorphic hardware design since the key factor governing power and energy consumption would be the average number of propagated spikes between neurons. Further, temporal spike encoding conveys more information than a rate-based code. Intuitively, this is driven by the following observation: Given an SNN computing framework operating in $n$ time steps, only $n$ different spike-streams can be encoded for rate-based encoding whereas temporal-domain encoding opens up the possibility of encoding $2^n$ different spike-streams. 

\begin{figure}
\centering
\includegraphics[width=0.48\textwidth]{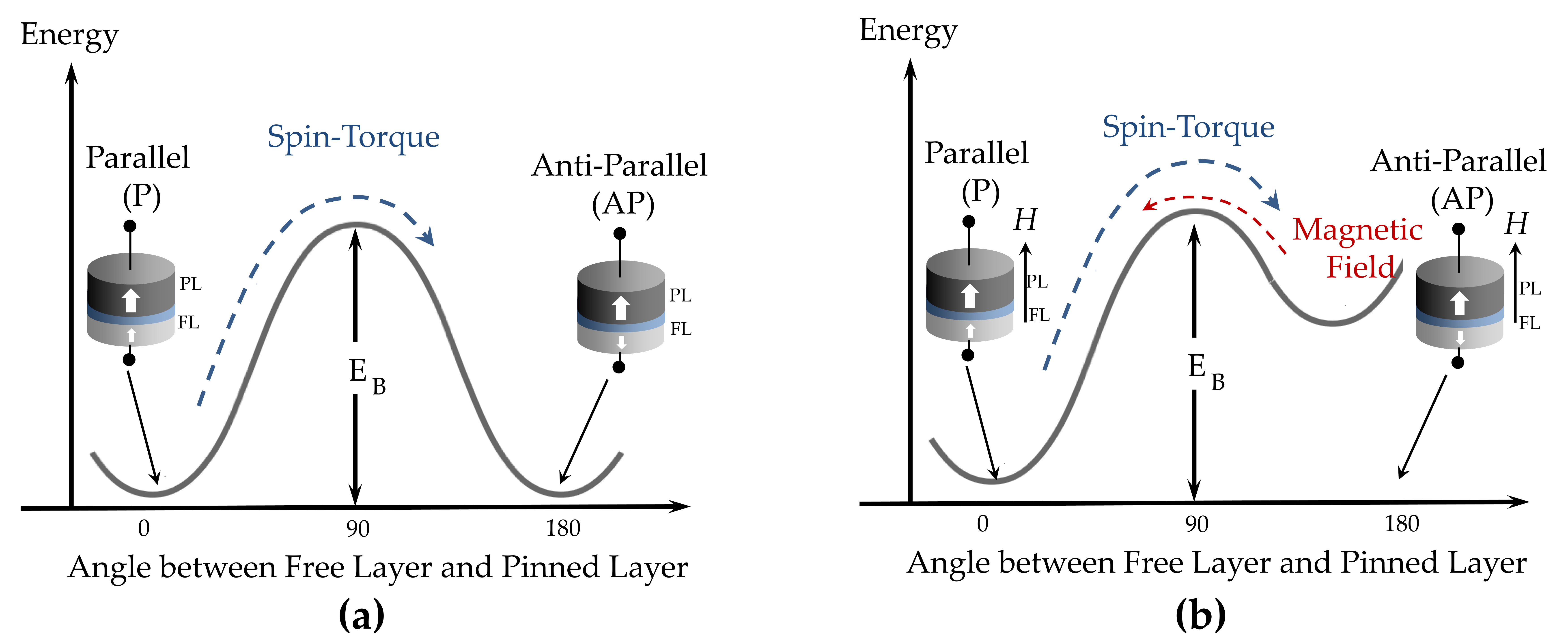}
\caption{\scriptsize{(a) In the absence of a magnetic field, the device spiking rate is modulated by the spin-torque generated by an external ``write" current. (b) Application of a magnetic field and spin-torque allows for independent control knobs for the individual device lifetimes.}}
\label{fig2}
\vspace{-2mm}
\end{figure}

The precise temporal variation of the output firing pattern of a stochastic spiking neuron is a function of the two lifetimes, $\tau_P$ and $\tau_{AP}$, in the P and AP states respectively. Note that the lifetimes represent the average magnitude in this text (the lifetime follows a Poisson probability distribution). In order to encode information temporally, precise control of the time-domain characteristics of the device is imperative, i.e. independent control of $\tau_P$ and $\tau_{AP}$ is required. This functionality can be conceptually envisioned in an MTJ stack by the application of an external magnetic field (see Fig. \ref{fig2}). In the absence of any magnetic field, the energy barrier in the P and AP states are equal when no external current is present. When a ``write" current is applied to the device, the AP state is favored. On the other hand, due to the application of the magnetic field, the P state energy barrier reduces. Hence, the magnitude of the external magnetic field and ``write" current are independent control knobs that can be used to modulate the device lifetimes, $\tau_P$ and $\tau_{AP}$, and hence the precise temporal code. Recent experiments on a CoFeB MTJ stack have demonstrated that such a control is indeed possible for a range of external fields and currents \cite{zink2018telegraphic}.  However, using an external tunable magnetic field to bias MTJ spiking neurons is not feasible from the scalability perspective for neuromorphic computing applications. Significant energy consumption would be required to generate the field. Additionally, since a tuned magnetic field is required for each specific neural magnet, this would limit the magnet spacing to avoid stray field effects. Taking inspiration from the core functionality and device physics explained above, we propose a modified MTJ structure that exploits magneto-electric effect to perform temporal encoding without the requirement of any external magnetic field.

Recent experiments on multilayered stacks consisting of a multi-ferroic material lying underneath a magnetic layer have revealed that a transverse magnetic field is induced in the nanomagnet lying on top due to the application of a voltage across the multi-ferroic material. This is attributed to Magneto-electric Effect (ME) \cite{fiebig2005revival}. ME generates from coupling between the spin polarization and the electric polarization of the material \cite{nikonov2014benchmarking}. The coupling is induced by Dzyaloshinskii-Moriya interaction, which occurs in crystal structures with certain symmetries \cite{dzyaloshinsky1958thermodynamic,moriya1960anisotropic}. ME has been observed in multiferroic materials such as $BiFeO_3$\cite{heron2014deterministic}. The applied electric field causes the displacement of bismuth ions inside $BiFeO_3$, followed by the rotation of oxygen octahedra. The shift of the ions results in the direction switching of ferroelectric polarization and magnetization. The switching of magnetization of $BiFeO_3$ acts as a bias to the contacting nanomagnet via exchange bias at the interface.
Note that this is just one possible route for realizing ME based devices. Other types of ME induced switching mechanisms \cite{cheng2018recent} can be potentially leveraged for our device design.

\begin{figure}
\centering
\includegraphics[width=0.44\textwidth]{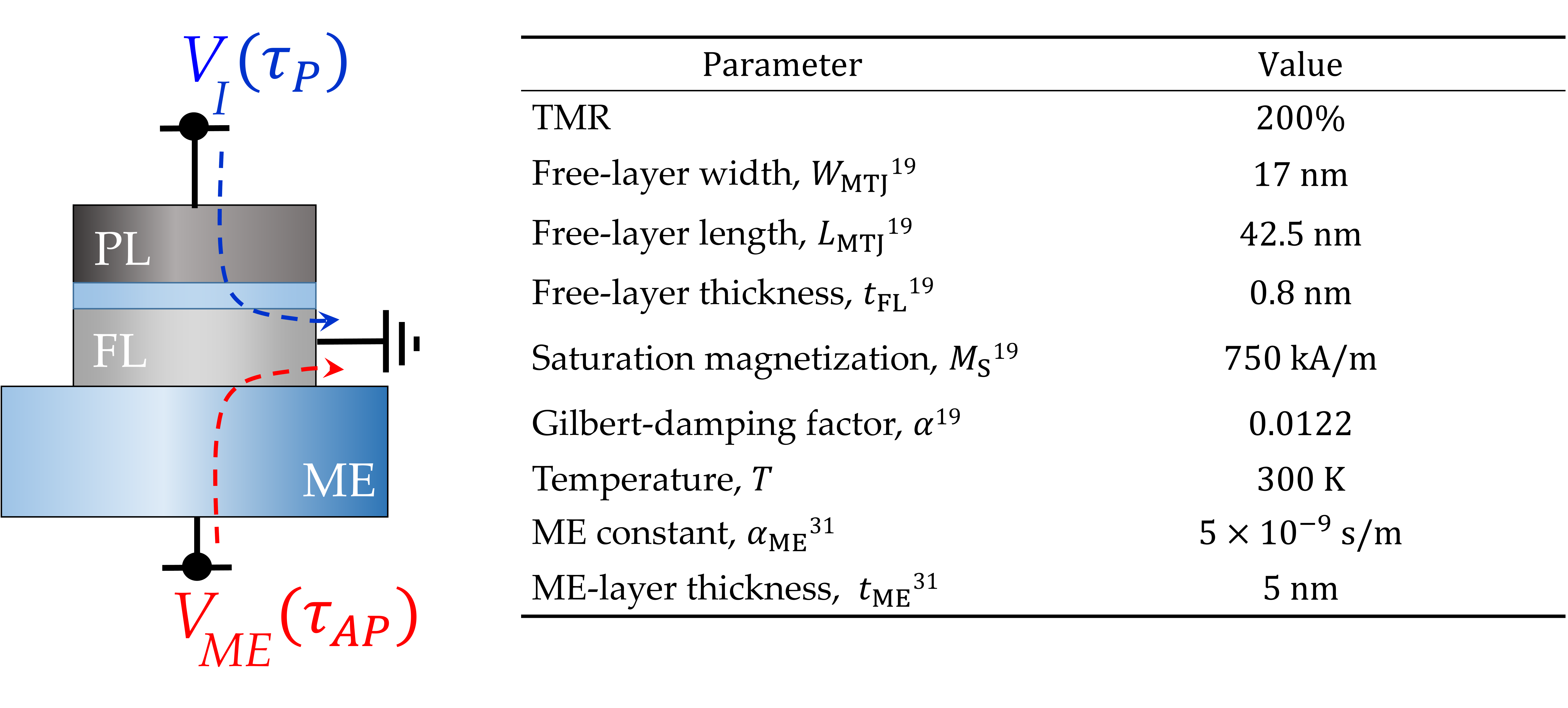}
\caption{\scriptsize{The magneto-electric device is driven by two independent inputs: (1) Voltage, $V_{ME}$, applied across the ME-oxide modulates lifetime $\tau_{AP}$, (2) Voltage, $V_{I}$, applied across the MTJ modulates $\tau_{P}$. Device simulation parameters are tabulated.}}
\label{fig3}
\vspace{-2mm}
\end{figure}

The probabilistic switching characteristics of an MTJ can be analyzed by Landau-Lifshitz-Gilbert (LLG) equation with additional term to account for the torque generated by the input spin current \cite{slonczewski1989conductance},
\begin{equation}
\label{llg}
\frac {d\widehat {\textbf {m}}} {dt} = -\gamma(\widehat {\textbf {m}} \times \textbf {H}_{eff})+ \alpha (\widehat {\textbf {m}} \times \frac {d\widehat {\textbf {m}}} {dt})+\frac{1}{qN_{s}} (\widehat {\textbf {m}} \times \textbf {I}_s \times \widehat {\textbf {m}})
\end{equation}
where, $\widehat {\textbf {m}}$ is the unit vector of FL magnetization, $\gamma= \frac {2 \mu _B \mu_0} {\hbar}$ is the gyromagnetic ratio for electron, $\alpha$ is Gilbert\textquoteright s damping ratio, $\textbf{H}_{eff}$ is the effective magnetic field including thermal noise \cite{scholz2001micromagnetic} and the shape anisotropy field for elliptic disks, $N_s=\frac{M_{s}V}{\mu_B}$ is the number of spins in free layer of volume $V$ ($M_{s}$ is saturation magnetization and $\mu_{B}$ is Bohr magneton), and $\textbf{I}_{s}$ is the input spin current. The spin current favors the AP state and is induced by an electric field applied across the MTJ stack, $I_{s}=V_{I}/R_{MTJ}$ ($R_{MTJ}$ is the resistance of the MTJ stack). ME effect is usually modelled by considering the effect of an external magnetic field acting on the magnet. The magnitude of the field is directly proportional to the applied voltage \cite{jaiswal2017mesl,nikonov2014benchmarking,chakraborty2018design}, with the proportionality factor being a material property. Note that this is agnostic to the underlying origin of ME and such a first-order relationship between applied voltage and induced magnetic field dependency have been extensively used for modelling and benchmarking magneto-electric devices \cite{jaiswal2017mesl,nikonov2014benchmarking,chakraborty2018design}. The applied magnetic field which favors P state due to the ME effect is given by,
\begin{equation}
    {\bf{H}}_{ME}=\left(0,0,\frac{1}{\mu_0}\alpha_{ME}\frac{V_{ME}}{t_{ME}}\right)
\end{equation}
where, $\alpha_{ME}$ is the ME constant, $t_{ME}$ is the thickness of ME layer and $V_{ME}$ is the voltage across the ME layer. It is worth noting here that the resistance of P state is smaller than that of AP state. Hence, the spin current is larger in the P state than in the AP state with the same applied $V_{I}$. Thus, a small variation in $V_{I}$ leads to a large change in spin current in P state. As a result, the $V_{I}$ ($V_{ME}$) control knob dominates $\tau_P$ ($\tau_{AP}$) variation. The asymmetric impact of each external voltage on $\tau_P$ and $\tau_{AP}$ enables independent control of the device lifetimes by applying two independent external control voltages. Note that the device can be still used to perform rate encoding by not utilizing the $V_{ME}$ control knob.

The magneto-electric effect can therefore be exploited to envision three-terminal device structures shown in Fig. \ref{fig3}. The device consists of an MTJ stack lying on top of an ME oxide layer (for instance, $BaTiO_3$ or $BiFeO_3$). Sufficient voltage ($V_{ME}$) applied across the ME oxide induces an effective magnetic field on the nanomagnet lying on top. On the other hand, the voltage applied across the MTJ, $V
_I$, controls the device lifetime $\tau_{P}$. Typical device simulation parameters for a $2k_{B}T$ barrier height magnet have been used from prior literature \cite{liyanagedera2017stochastic,chakraborty2018design} and are tabulated in Fig. \ref{fig3}. 

The main distinguishing factors in our neuromimetic ME-MTJ design are as follows: \textbf{(i)} ME-MTJs have typically been considered to be switched by applying a voltage across the ME oxide \cite{jaiswal2017proposal}. Here, we propose to use two independent inputs ($V_{ME}$ and $V_{I}$). While the voltage applied across the ME oxide will produce the effect of an applied magnetic field (thereby modulating $\tau_{AP}$), the external input current will be used to control $\tau_{P}$. Independent control of these two parameters will enable us to implement a stochastic nanoelectronic spiking neuron functionality that inherently performs temporal domain encoding of information, as explained in the previous section. \textbf{(ii)} Most of the work on ME-MTJs are catered for usage of these devices in logic and memory applications \cite{manipatruni2018beyond,nikonov2014benchmarking,jaiswal2017mesl,jaiswal2017energy,chakraborty2018design}. Our proposal involves utilizing the ME for enabling neuromorphic applications, and in particular, for temporal-encoding of stochastic SNNs.
\begin{figure*}[t!]
    \centering
    \includegraphics[width=0.62\textwidth]{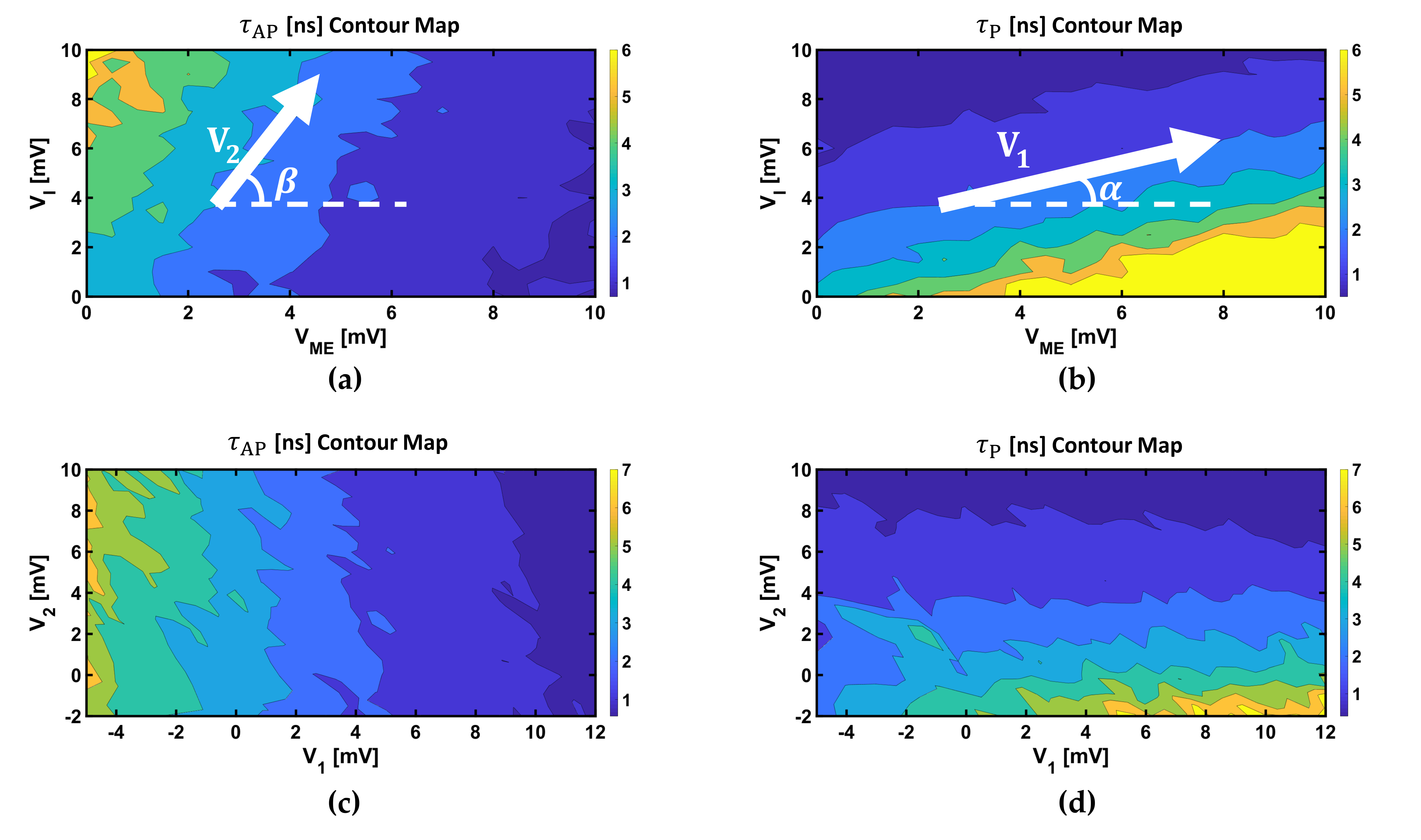}
    \caption{\scriptsize{(a-b) Contour map of $\tau_P$ and $\tau_{AP}$ versus $V_{I}$ and $V_{ME}$, (c-d) Contour map of $\tau_P$ and $\tau_{AP}$ versus $V_{1}$ and $V_{2}$}.}
    \label{contour}
    \vspace{-6mm}
\end{figure*}

Next, we characterize the device operation by varying the two external input voltages and measuring the average device lifetimes. It is worth noting here that from a system development perspective, the neurons will be interfaced with synaptic devices. Hence, achieving truly independent control of $\tau_P$ and $\tau_{AP}$ over a wide operating range of $V_{I}$ and $V_{ME}$ is crucial. We define a set of $k$ factors to evaluate the impact of the two external bias signals on $\tau$:
\begin{equation}
    k_{AP(P),ME(I)}= \frac{\partial \tau_{AP(P)}}{\partial V_{ME(I)}}  \label{kfactors}
\end{equation}
The value of $k$ depicts the amount of change in $\tau$ induced by a unit change in one of the bias ($V_{ME}$ or $V_I$) with the other bias fixed. The total change of $\tau$ is expressed as
\begin{equation}
    \Delta\tau_{AP(P)}=k_{AP(P),ME}\Delta V_{ME}+k_{AP(P),I}\Delta V_I
    \label{dtau}
\end{equation}

To realize the independent control, variation of $\tau$ in one state should be dominated by only one of the bias, leading to the condition:
\begin{equation}
\left\vert\frac{k_{AP,ME}}{k_{AP,I}}\right\vert\gg1,     \left\vert\frac{k_{P,I}}{k_{P,ME}}\right\vert\gg1    
    \label{condition1}
\end{equation}

Eq.(\ref{condition1}) implies that $\tau_{AP}$ is dominated by $V_{ME}$ and $\tau_P$ is dominated by $V_I$. It should be noted that the ratios in Eq.(\ref{condition1}) are related to the slope of contour lines of $\tau_{AP}$ and $\tau_P$ for varying $V_{ME}$ and $V_I$. Fig.~\ref{contour} depicts the contour map of $\tau$ in the two states. The contour lines in P state have a small slope (note that $\left\vert\frac{k_{P,I}}{k_{P,ME}}\right\vert$ is the reverse of the slope), indicating that the spin current has a dominant control on $\tau_{P}$, while the slope of contour lines in AP state is large, which implies that ME is the leading control factor. 

However, in order to achieve truly independent control, the change of the dominant bias in one state should not make a prominent change on $\tau$ of the other state. For example, since $V_{ME}$ dominates $\tau_{AP}$, $\Delta\tau_{AP}$ induced by $\Delta V_{ME}$ need to be much larger than $\Delta\tau_P$ induced simultaneously. As a result, another condition for the independent control is
\begin{equation}
\left\vert\frac{k_{AP,ME}}{k_{P,ME}}\right\vert\gg1,    \left\vert\frac{k_{P,I}}{k_{AP,I}}\right\vert\gg1 
    \label{condition2}
\end{equation}
Eq.(\ref{condition2}) indicates $V_{ME}$ has a much larger impact on $\tau_{AP}$ than on $\tau_P$, and $V_I$ has a larger impact on $\tau_P$ than on $\tau_{AP}$. According to the definition of $k$, larger $k$ values result in more rapid change of $\tau$, leading to denser contour lines in the contour map. From this perspective, Eq.(\ref{condition2}) states that our device operating region has to restricted in an area where the contour lines should be much denser (the spacing between adjacent contour lines is smaller) in AP state going along the $V_{ME}$ axis, and concurrently the contour lines are much denser in P state going along the $V_I$ axis. As is shown in Fig.~\ref{contour}, in the map of AP state, contour lines are denser in the top-left while in the P state, the denser area is in the bottom-right portion of the plot. This opposite nature of $k$ factor variation severely limits the operating region of the device toward the middle diagonal region of the plot to compromise between the restrictions imposed by Eq. (\ref{condition2}).

Interestingly, we observe that although the contour lines are not strictly horizontal or vertical, the slope of the lines is approximately constant throughout the entire range. Hence, to remove the limitation and expand the device operating region, one can introduce a set of new basis signals in the direction of the contour lines in Fig.~\ref{contour}. No unwanted $\Delta\tau$ will be induced as $\tau$ is fixed along the contour lines. The new basis signals are given by,
\begin{equation}
\begin{pmatrix}
V_1 \\
V_2
\end{pmatrix} = \begin{pmatrix}
\cos \alpha & \sin \alpha \\
\cos \beta & \sin \beta 
\end{pmatrix} \begin{pmatrix}
V_{ME} \\
V_I
\end{pmatrix}
\end{equation}
where $\alpha$, $\beta$ are shown in Fig.~\ref{contour}(a-b). The contour map based on the new basis $V_{1(2)}$ is plotted in Fig. \ref{contour}(c-d). The neuron functionality can be now conceptualized as being driven  by external inputs $V_1$ and $V_2$. The actual inputs to the device, $V_{ME}$ and $V_I$, is a linear combination of the two external inputs, $V_1$ and $V_2$, which can be easily implemented by voltage divider circuits. From a network perspective, these signals would be determined by current flowing though synaptic devices \cite{sengupta2016proposal}. We would like to note here that such a simple transformation is made possible due to the constant slope of contour lines throughout the plot. As observed in Fig. \ref{contour}(c-d), the contour lines are approximately horizontal/vertical, thereby realizing independent control of device lifetimes over the entire operating region.



Our proposed magnetoelectric device is an addition to the toolset of superparamagnetic devices enabling the recent wave of unconventional probabilistic computing scenarios \cite{roy2018perspective,camsari2019p}. Note that the conclusions presented in this article is not specific to the magnet barrier height. In summary, such magneto-electric devices inherently perform temporal encoding in the functionality of a stochastic spiking neural unit and bears the potential of enabling an alternative sparse, low-latency neuromorphic computing paradigm. Future work will explore algorithms that can leverage such fine-grained temporal manipulation of the magnetization of stochastic nano-neurons.


%

\end{document}